# Sub-meV Linewidths in Polarized Low-Temperature Photoluminescence of 2D PbS Nanoplatelets

*Pengji Li[‡,1], Leon Biesterfeld[‡,2,3,4], Lars Klepzig[2,4], Jingzhong Yang[1], Huu Thoai Ngo[5],*

*Ahmed Addad[6], Tom N. Rakow[1], Ruolin Guan[1], Eddy P. Rugeramigabo[1],*

*Ivan Zaluzhnyy[7], Frank Schreiber[7],*

*Louis Biadala[\*,5], Jannika Lauth[\*,2,3,4,5,8], Michael Zopf[\*,1,8]*

[1]Institute of Solid State Physics, Leibniz University Hannover, Appelstraße 2, D-30167 Hannover, Germany.

[2]Cluster of Excellence PhoenixD (Photonics, Optics, and Engineering – Innovation Across Disciplines), Welfengarten 1A, D-30167 Hannover, Germany.

[3]Institute of Physical and Theoretical Chemistry, Eberhard Karls University of Tübingen, Auf der Morgenstelle 18, D-72076 Tübingen, Germany.

[4]Institute of Physical Chemistry and Electrochemistry, Leibniz University Hannover, Callinstr. 3A, D-30167 Hannover, Germany.

[5]Université de Lille, CNRS, Centrale Lille, Université Polytechnique Hauts-de-France, Junia-ISEN, UMR 8520 - IEMN, F-59000 Lille, France.

[6]Université Lille, CNRS, INRAE, Centrale Lille, UMR 8207 − UMET- Unité Matériaux et Transformations, F-59000 Lille, France.

[7]Institute of Applied Physics, Eberhard Karls University of Tübingen, Auf der Morgenstelle 10, D-72076, Tübingen, Germany.

[8]Laboratory of Nano and Quantum Engineering, Leibniz University Hannover, Schneiderberg 39, D-30167 Hannover, Germany.

ABSTRACT

Colloidal semiconductor nanocrystals are promising materials for classical and quantum light sources due to their efficient photoluminescence (PL) and versatile 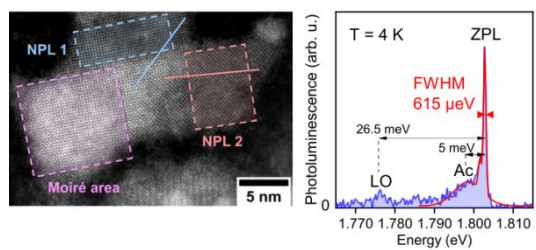 chemistry. While visible emitters are well-established, the pursuit for excellent (near-)infrared sources continues. One notable candidate in this regard are photoluminescent two-dimensional (2D) PbS nanoplatelets (NPLs) exhibiting excitonic emission at 720 nm (1.7 eV) directly tying to the typical emission range limit of CdSe NPLs. Here, we present the first comprehensive analysis of low-temperature PL from this material class. Ultrathin 2D PbS NPLs exhibit high crystallinity confirmed by scanning transmission electron microscopy, and reveal Moiré patterns in overlapping NPLs. At 4 K, we observe unique PL features in single PbS NPLs, including narrow zero-phonon lines with linewidths down to 0.6 meV and a linear degree of polarization up to 90%. Time-resolved measurements identify trions as the dominant emission source with a 2.3 ns decay time. Sub-meV spectral diffusion and no immanent blinking over minutes is observed, as well as discrete spectral jumps without memory effects. These findings advance the understanding and underpin the potential of colloidal PbS NPLs for optical and quantum technologies.

**Introduction**

Colloidal semiconductor nanocrystals are being extensively studied for their use as classical and quantum light sources due to their optical properties dominated by size quantization.[1–3] Key requirements for their application include photo-stable PL with high quantum yields, short radiative lifetimes, low spectral broadening and diffusion, as well as scalable fabrication. At UV to visible wavelengths, cadmium chalcogenide CdX (X = S, Se, and Te) NPLs and heterostructures are known for their excellent optical properties. In particular, CdSe-based NPL systems exhibit narrow linewidths below 40 meV at room temperature,[4,5] between 80 % and up to unity PL quantum yield,[6–8] fast radiative decay in the nanosecond range,[4,9] and highly directional PL.[10] In recent years, colloidal lead halide perovskite nanocrystals have emerged as emitters with efficient (quantum yield over 96 %),[11] narrow (12 – 42 nm) and rapidly decaying (1 – 29 ns) room-temperature PL at visible wavelengths,[12] as well as single photon emission.[13–15] However, materials with similar characteristics at (near-)infrared (NIR) wavelengths are yet highly sought for, in particular for photonic quantum communication applications.[16–18] Potential candidates include Ag-doped CdSe NPLs,[19] HgTe NPLs[20] or InAs/CdSe core-shell nanocrystals.[21] Another promising material class are lead chalcogenide PbX (X = S, Se, Te) QDs[22–24], NPLs[25–27] and related heterostructures[28–30]: For instance, Krishnamurthy *et al.* demonstrated single spherical PbS/CdS QDs emitting in the telecom O-band (near 0.95 eV) at room-temperature, featuring photon antibunching and an average linewidth of 89.5 meV.[29] In a similar system and at T = 4K, Hu *et al.* reported on bleaching-free PL at around 1.0 eV and mean intrinsic PL linewidth of 16.4 meV, featuring asymmetric line shapes caused by the coupling of excitons to optical and acoustic phonon modes.[28] In both cases, the broad linewidths (compared to their II-VI analogues, such as CdSe QDs) are a result of a 64-fold degenerated band-edge exciton in PbX QDs that splits into multiple energetically similar transitions, resulting in intrinsic PL broadening.[28,31,32] A closely related, yet unexplored system at the single particle level are photoluminescent 2D PbS NPLs. Manteiga Vázquez *et al.* developed a synthesis of rock salt cubic-structured PbS NPLs exhibiting a PL quantum yield of up to 19.4 % for PL at 1.7 eV (720 nm) upon surface passivation with $CdCl_2$.[25] This strongly enhanced emission efficiency provides the opportunity to investigate the excitonic emission properties of 2D PbS NPLs at the individual emitter level and exploring their electronic structure, phonon interactions and spectral characteristics at cryogenic temperatures.

Our findings provide the first in-depth optical study of individual PbS NPLs at cryogenic temperatures. Highly polarized emissions at around 1.82 eV (677.6 nm) with sub-meV linewidths are observed at T = 4K, accompanied by an acoustic phonon sideband. Time-resolved and excitation power dependent measurements reveal trion states as the dominant cause of PL. The emissions exhibit exceptional spectral stability with sub-meV spectral diffusion and are blinking-free over several minutes, promoting the potential of 2D PbS NPLs for reaching near-infrared optoelectronic applications.

**Results and discussion**

NIR emitting colloidal PbS NPLs passivated with $CdCl_2$ were synthesized by a method described by Manteiga Vázquez *et al.*[25] Figure 1a shows an overview TEM image of PbS NPLs resembling a rectangular shape with average lateral dimensions of (16.0 ± 1.6) x (9.2 ± 1.2) $nm^2$ and a corresponding aspect ratio of 1.7:1 (see Figure S1 for an additional overview image and the corresponding size histogram). Figure 1b depicts a HR-HAADF-STEM image of two overlapping PbS NPLs (the corresponding FFT patterns of the highlighted crystal areas are shown in Figure S2a). The individual PbS NPLs are highly crystalline and exhibit the cubic rock salt structure (space group $Fm\bar{3}m$) expected for 2D PbS nanosheets (NSs) and NPLs[25,33,34] with the characteristic lattice spacings of 2.9 Å (200) and 2.1 Å (220) (PDF card 01-078-1900). Notably, no diffraction peaks of an orthorhombic PbS phase (interplane distances of 2.8 Å and 2.05 Å)[35] are evident from the FFT patterns (see Figure S2). The ultrathin 2D geometry and crystalline nature of the NPLs is underpinned by the formation of a pronounced Moiré pattern in the overlapping area of the two differently oriented diffracting crystals (see Figure S2b, c for additional examples). Although not directly related to the in-depth optical studies in this work, the formation of randomly oriented Moiré patterns suggests that the small $CdCl_2$ ligands used as X- and Z-type ligands in a post-synthetic surface passivation step, allow for a quasi-direct contact between some PbS NPLs (in contrast to typical bulky organic surfactants such as oleic acid, which lead to further spatially separated NPLs, see also Figure S3 for grazing-incidence wide-angle X-ray scattering data of PbS NPLs and further discussion).[26,36] While twistronics are very thoroughly researched for van der Waals materials,[37] Moiré superlattices based on metavalently bound materials such as PbS have only recently been accessed by Wang *et al. via* an aqueous synthesis route with readily removable ligands. We assume that metal halide passivation can yield similar formations while at the same time enhancing the optical properties of the PbS NPLs synthesized in organic medium.[36] Figure 1c depicts the optical characteristics of the PbS NPL ensemble in colloidal solution at room-temperature, which exhibit an excitonic absorption feature at 1.96 eV and associated NIR PL at 1.70 eV with a rather broad FWHM of 264 meV (Figure 1c). To gain further insight into the optical, structural and electronic properties of PbS NPLs at the single NPL level, we perform PL measurements at cryogenic temperatures (see SI, section A).

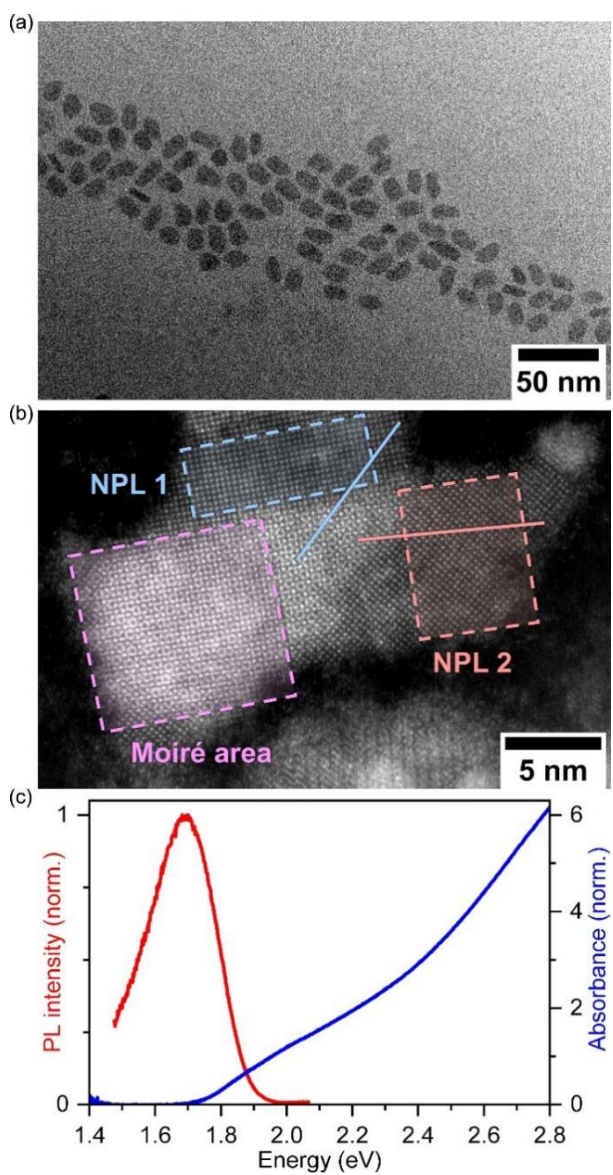

**Figure 1.** (a) Overview TEM image of rectangular PbS NPLs exhibiting average lateral dimensions of (16.0 ± 1.6) x (9.2 ± 1.2) nm$^2$. (b) HR-HAADF-STEM image of two overlapping PbS NPLs forming a Moiré pattern. The formation of the interference pattern emphasizes the ultrathin 2D geometry of the PbS NPLs. (c) Ensemble room temperature absorbance and PL spectrum of PbS NPLs (in colloidal solution), exhibiting excitonic absorption at 1.96 eV and NIR PL at 1.70 eV.

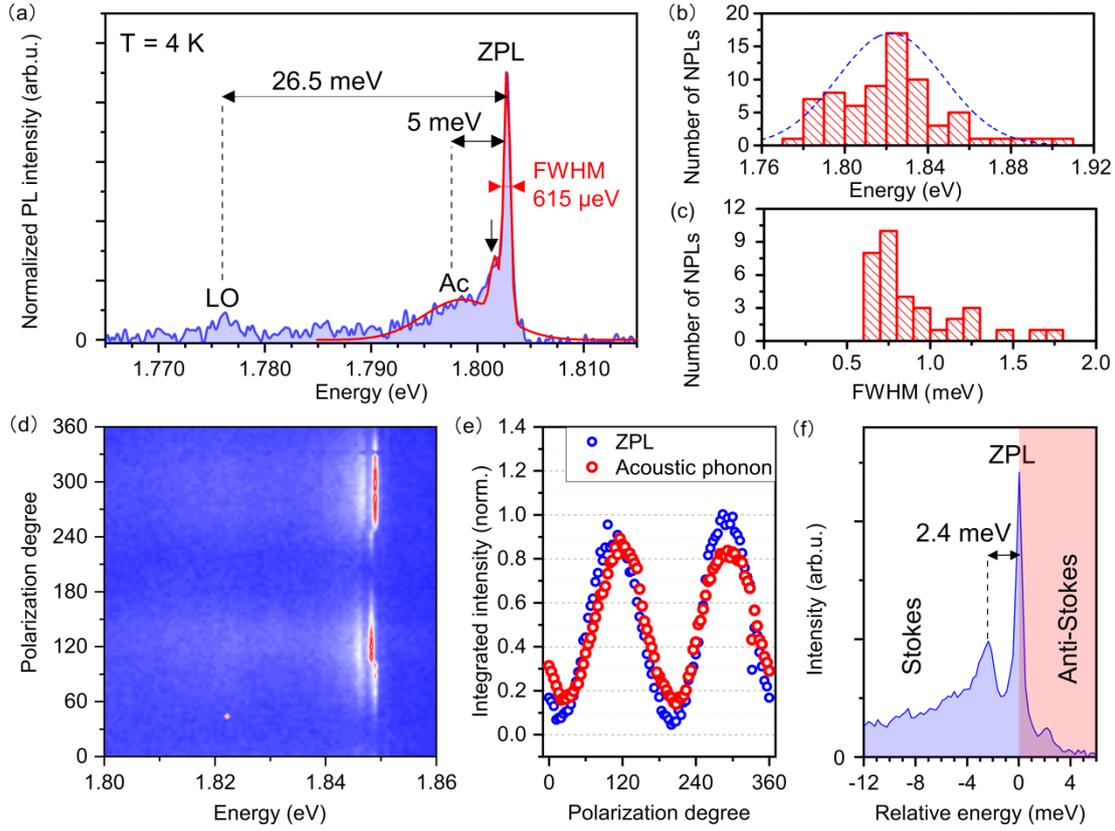

**Figure 2.** PL of single PbS NPLs at T = 4 K. (a) Micro-PL spectrum of a single PbS NPL (1s exposure time) featuring sub-meV emission and a red-shifted phonon sideband with LO-phonon replicas (LO) and acoustic phonon (Ac) contributions. (b) Distribution of the emission energy centered around 1.82 eV and (c) FWHM of the narrowband part of single PbS NPL emissions, obtained by measuring 71 individual NPLs. 74% of the measured emission lines exhibit sub-meV linewidths. (d) Polarization-dependent PL measurement of a single PbS NPL and (e) the respective normalized PL intensity obtained in different spectral ranges as a function of the linear polarization angle of the emitted light. The blue dots represent the ZPL ($\delta$ = 0.90), the red dots correspond to the acoustic phonon sideband ($\delta$ = 0.71). (f) Normalized sum of the 91 PL spectra from the measurement in (d) and illustration of Stokes and anti-Stokes PL of the acoustic phonon sidebands of a single PbS NPL.

Figure 2a shows a representative PL spectrum of individual PbS NPLs at T = 4 K. In marked contrast with typical PL spectra of PbS nanocrystals for which single broad (≥ 8 meV) emission lines are observed,[28] the PL spectra of PbS NPLs consist of a narrow, resolution-limited, zero phonon line (ZPL) together with phonon sidebands at energies of 26.5 meV and 5 meV that we attribute to optical (LO) and acoustic phonons (Ac), respectively[28,38] (see Table S1 for the fitting results, Figures S5 and S6 for more spectra of individual NPLs). The statistics on more than 70 individual NPLs show that the emission energies of the ZPL (Figures 2b) is centered around 1.82 ± 0.06 eV, which clearly indicates a high level of uniformity in the thickness and the lateral dimensions of the PbS NPLs. Moreover, Figure 2c shows that the NPL emission linewidth does not exceed 2 meV and that more than 74 % of the studied PbS NPLs exhibit a sub-meV linewidth. The detected linewidths go down to 0.6 meV, approaching the resolution limit of the spectrometer. This strongly contrasts with the sharpest emission linewidth measured on individual PbS nanocrystals ($\geq$ 8 meV) [28] for which the line broadening stems from the exciton fine structure, [28] the intervalley and the exciton-phonon coupling effects,[39] and the spectral diffusion.[40] Therefore, the record sharpness of the PbS NPLs studied here points to i) an absence of spectral diffusion, ii) a reduced exciton-phonon coupling and/or iii) a different excitonic origin of the emission. Strikingly, we observe an additional discrete peak (arrow in Figure 2a) close to the ZPL on the PL spectra. A detailed analysis of the PL spectra around the ZPL (conducted on another NPL, Figure 2f) unveils that this discrete peak appears in both, the Stokes and anti-Stokes part of the PL spectra. Such a well-defined peak observed around 2.4 meV for most of the NPLs is most likely stemming from the thickness breathing mode (which would be 2.4 meV for a 1.6 nm thick PbS NPL, see Figures S7 for TEM images of PbS NPLs exhibiting the thickness of 1-2 nm).[41,42] This feature, previously observed on PL spectra of individual CdSe[43] and InP[44] NCs at cryogenic temperatures, is the fingerprint of confined acoustic phonon modes at about 10 K, which corresponds to the base temperature in cold-finger cryostats for such sample preparation.

Notwithstanding unprecedented sharpness of their emission lines, PbS NPLs exhibit striking linear polarization properties (Figure 2d), which are analyzed by utilizing a rotating half-wave plate followed by a polarizer. In Figure 2e, integrated intensities of the ZPL and the acoustic phonon sideband for the various polarization angles are reported. From the angle dependent PL intensity, we evaluate the polarization degree, $\delta$, as $\delta = (I_{max}-I_{min}) / (I_{max}+I_{min})$ where $I_{max}$ ($I_{min}$) are the maximum (minimum) PL intensities. The ZPL displays a high polarization degree $\delta = 0.90$ (for similar data from a second PbS NPL, we refer to Figure S8). The acoustic phonon sideband (from -1 meV to -8 meV in Figure 2f) shows a slightly lower polarization degree of $\delta = 0.71$, while maintaining the same polarization angle as the ZPL.

The emission polarization of single NPLs is influenced by their aspect ratio,[31] as well as the orientation of the NPL with respect to the substrate. A dipole orientation in the plane of the substrate will show a maximum polarization degree, whereas dipoles with orthogonal orientations are expected to appear as unpolarized emission (details in the SI, section B, Figure S8). The high degree of linear polarization in PbS NPLs indicates that the excitonic transition in individual PbS NPLs exhibits a polarization component, attributed to a linear 1D or 2D dipole. Furthermore, the alignment of the dipole is nearly ideal to the substrate plane (similar to observations shown in Figure 1). The aspect ratio is PbS NPLs is approx. 2, leading to anisotropic lateral electronic confinement and therefore contributing to the enhanced degree of observed polarization. It is important to note that the degree of polarization may depend on further factors that we do not study in detail here, such as selection rules of the allowed excitonic states and the respective oscillator strengths or possible effects of absorption polarization, by which the recombination of specific exciton types can be favored depending on the excitation energy and the energy spectrum of the NPL.

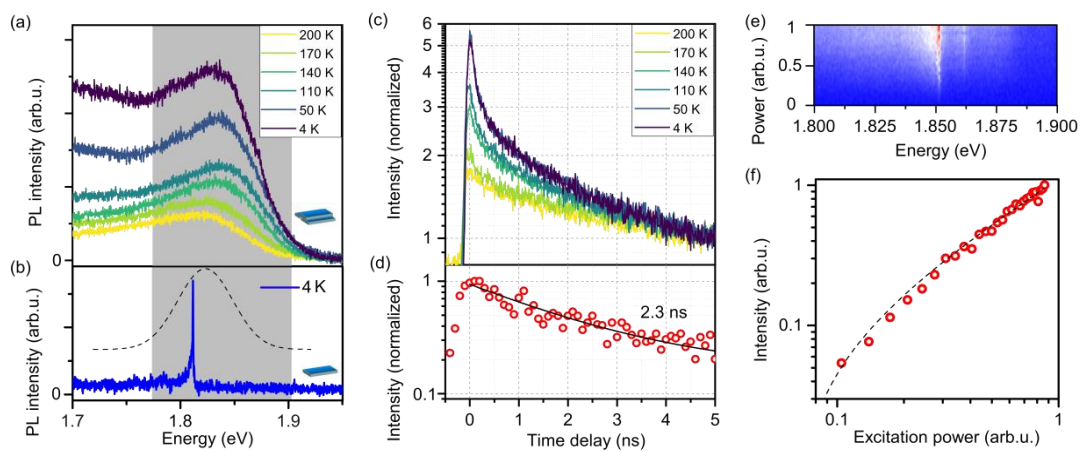

**Figure 3.** (a) Temperature-dependent PL spectra of an ensemble of PbS NPLs and (b) PL spectrum of a single PbS NPL at 4 K. The gray shaded area indicates the spectral range selected for lifetime measurements. The black dashed line represents the energy distribution of emissions from single PbS NPLs at 4 K in Figure 2b. (c) Fluorescence lifetime measurements of an ensemble of PbS NPLs, normalization is performed at a 'long' time scale (5 ns in this case) to emphasize the fast component and its strong temperature dependence. (d) Fluorescence lifetime measurement of a single PbS NPL. A mono-exponential decay model is applied (solid, black line). (e) Excitation laser power dependent PL spectra of a single PbS NPL and (f) the respective normalized integrated PL intensity of the ZPL showing a linear increase with the excitation power.

Figure 3 shows the comparison of a PbS NPL ensemble with single NPL PL to study the origin and excitonic nature of the emission in more detail. Figure 3a includes the temperature-dependent PL spectra of a PbS NPL ensemble (Figure S9 shows the data with an even more precise stepwise increase in temperature). We find an increasing slope of the high-energy PL edge with decreasing temperature, indicating an increased relative intensity and a decreased spectral width of the bandgap-associated emission.

The PL spectrum of a single NPL at 4 K in Figure 3b, as well as the distribution of emission energies discussed in Figure. 2b, shows good overlap with the emission edge of the ensemble emission at T = 4 K in Figure 3a.

The spectral region of the emission is then filtered (shaded gray area in Figure 3a, b) and detected with an avalanche photodiode to perform PL decay measurements for temperatures from 4 K to 200 K (Figure 3c). The decay dynamics occur on two distinct timescales: a slow component ($\tau_1 \geq 6$ ns) and a fast component ($\tau_2$ in the sub-nanosecond range), with detailed data provided in Figures S9. As the temperature increases to 160 K, we observe a lengthening of the fast decay component from 300 ps to 750 ps. The fast decay component, $\tau_2$, remains stable between 10 K and 70 K, with most changes occurring above 70 K, which is consistent with the findings of Canneson *et al.*[45] on trion emission in CsPbBr$_3$ quantum dots. At temperatures of 170 K, the fast decay component disappears. This vanishing of the fast component could be explained by the thermal energy overcoming the binding energy of the excitonic complex, allowing us to estimate the trion binding energy to be ~14.7 meV, which is close to the theoretical values reported in the literature for PbS NPLs of similar dimensions.[46] (similar temperature-dependent results for PbS NPLs ensembles can be observed in Figure S10 and S11). These dynamics stand compared to the PL decay of PbS nanocrystals, which typically occurs on a microsecond timescale (see Figure S12 for PL lifetime measurement of ensemble PbS NPLs at RT).[47,48] We attribute the presence of the fast component to the generation of higher order excitonic complexes (such as trions), additionally favored by a pile-up effect due to the high repetition rate of the pulsed laser of 82 MHz used for photoexcitation in the temperature-dependent measurements.

The PL decay of a single PbS NPL at 4 K is shown in Figure 3d, and data is best modelled with a mono-exponential decay, even though it cannot be ruled out completely that a smaller fraction with a slower decay component is present also in this case. A lifetime of 2.3 ns is extracted, which is in the realm of lifetimes reported for CdSe[49] and InP[50] NCs.

A further tool to investigate the excitonic origin of NPLs and nanocrystals are excitation power dependent measurements.[51,52] We perform these measurements under CW excitation with results shown in Figure 3e. A linear increase of the integrated PL intensity of the ZPL (Figure 3f) and the lack of additional spectral features with growing excitation power rule out multi-excitonic emissions as well as recombination processes that are independent of the excitation power (such as those involving trap or defect states).

By considering all observations up to this point, the question arises whether the narrow-band emission in single PbS NPLs can either be attributed to the presence of neutral excitons or trions, i.e. an exciton in presence of an additional unpaired charge carrier. Owing to their specific spin structure,[53] optical properties of trions (positive and negative) include rather temperature-independent emission characteristics.[54] This results in trion emission having a rapid initial decay phase, which is indicative of swift radiative recombination. This phase is faster than the decay of dark excitons, yet slower than that of bright excitons.[54] The unique dynamics of neutral excitons, which are susceptible to temperature due to the interplay between optically inactive (dark) and active (bright) states, are in stark contrast with the more temperature-stable behavior of trions. However, the proportion of trion transitions is sensitive to temperature, particularly in relation to their binding energy. When thermal energy surpasses the trion binding energy, the trion emission effectively vanishes. This aligns well with observations from the PbS NPL ensemble PL decay as a function of temperature, which demonstrates that, while the rate of fast PL decay remains constant, its relative weight diminishes with an increase in temperature. Based on these factors, we assume the fast decay of the narrowband emission of single PbS NPLs at 4 K is initiated by a trion transition. The sign of the trion does not affect the conclusion but further studies on the magneto-optical properties of the trion will aim at unveiling the sign of the trion and their spin dynamics.

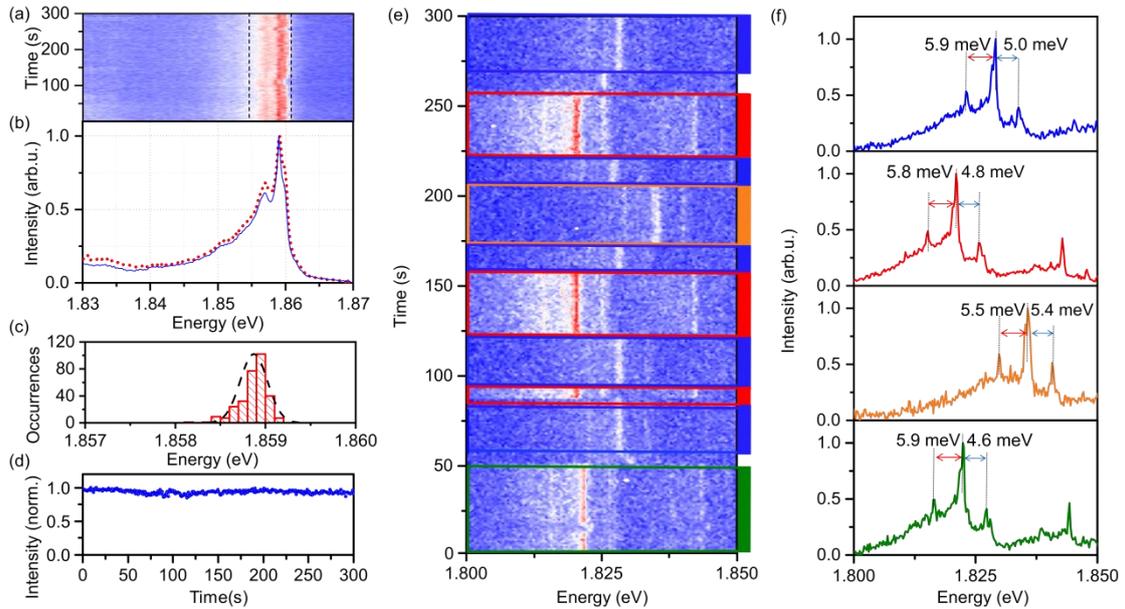

**Figure 4.** Spectral and temporal dynamics of single PbS NPLs. (a) PL time traces of a representative PbS NPL, exhibiting minimal spectral diffusion and blinking. (b) Normalized sum (red, dashed line) of the 300 PL spectra from (a). The blue line features the normalized sum of spectra which is corrected for spectral diffusion by rescaling the x-axis of each spectrum to the emission energy of the strongest emission peak. (c) Distribution of the central emission energy of the emission (obtained by fitting each spectrum in (a)) and (d) intensity time trace from the emission in (a) within the integration range shown by the dashed line. (e) PL time trace of a single PbS NPL, revealing combined spectral diffusion and blinking with discrete jumps. (f) Normalized sum of spectra for the four distinct emission states observed in (e), with the spectrums' color corresponding to the selected ranges in (e).

To characterize the spectro-temporal dynamics in PbS NPLs, we analyze PL time traces of individual NPLs at a cryostat temperature of 4 K. These dynamics are key to understanding the influence of the nanomaterials' surroundings on their photophysical properties. Figure 4a shows 300 consecutively recorded spectra with 1 s exposure time each, exhibiting a highly stable emission of PbS NPLs. All recorded spectra are summed up and normalized to obtain a time-integrated spectrum over 300 s (dashed red line in Figure 4b). A second approach addresses the spectral diffusion by rescaling the x-axis of each individual spectrum to match the emission energy of the strongest emission peak, followed by summing up these adjusted spectra and normalizing them (solid blue line in Figure 4b). The excellent overlap between the normalized sum of spectra and the spectral diffusion-compensated normalized sum of spectra underpins that the effect of spectral diffusion in the single PbS NPL emission is almost negligible over 5 min (see Figure S13 and S14 for more time trace measurements of individual PbS NPLs). The distribution of the central emission energy, derived from applying a Gaussian line shape to each spectrum, is presented in Figure 4c and reveals sub-meV (FWHM of 0.4 meV) spectral diffusion. The reduced fast spectral dynamics, which typically occur due to the Stark effect induced by trapped surface charges, points to a low density of surface traps. The application of $CdCl_2$ in a post-synthetic step for surface passivation has been identified as an effective strategy for addressing dangling bonds *via* X- and Z-type binding to $Pb^{2+}$ and $S^{2-}$ surface sites, respectively, and contributes to a reduced trap state density.[25,26] Mid-gap trap states are also expected to be of low level of significance due to the high crystallinity and well-balanced stoichiometry (Figure 1 and also Manteiga *et. al.*[20]), as well as an expected robustness of PbS NPLs against off-stoichiometry predicted in *ab-initio* simulations.[55] Figure 4d shows the time-dependent intensity of the emitted signal shown in Figure 4a, within the spectral region indicated with dashed lines. A stable emission is observed over long times with no clear traces of blinking and only slow, low-magnitude drifts of the intensity, which is yet another argument for a low trap-state density in PbS NPLs. While the low spectral diffusion is a common feature of the studied NPLs, the spectral characteristics can vary considerably. Figure 4e illustrates the spectral evolution of another PbS NPL, which displays several discrete spectral jumps over time. Four distinct spectral positions were identified, and for each of these we show the normalized sum of the respective spectra in Figure 4f by using the same color. Despite the differences in spectral position, the spectra are almost identical: The ZPL as well as observed Stokes and anti-Stokes PL features with phonon energies of around 5-6 meV do not change significantly. Another noteworthy feature is the lack of a memory effect[56] at the spectral jumps, which indicates that the spectral positions correspond to discrete localized states that are not correlated or affected by the previous state of the NPL. These observations could be consistent with a trion emission experiencing four different Coulomb environments (e.g., four localized trapping sites) where the spectral positions are determined by the quantum-confined Stark effect induced by a charge carrier trapping. The absence of spectral diffusion (see Figure S15) at each spectral position indicates that the charges remain strongly localized as hopping charges would cause Stark effect fluctuations.

**Conclusion**

In conclusion, we have synthesized highly crystalline, ultrathin 2D PbS NPLs with $CdCl_2$ ligands used for surface passivation. A comprehensive analysis of the PL properties of these PbS NPLs is conducted at cryogenic temperatures. The results reveal that single PbS NPLs exhibit strong and linearly polarized emission at 4 K, showcasing sub-meV linewidths significantly narrower than those observed in spherical nanocrystals of similar materials. These findings highlight the unique optical properties conferred by the 2D geometry of the PbS NPLs. Time-resolved PL measurements confirm that the narrow emission originate from trions. The trion state in PbS NPLs demonstrates stable emission with minimal spectral diffusion and the absence of blinking over minutes. Additionally, PbS NPLs exhibit new spectral diffusion characteristics, which lack a memory effect. Our findings not only advance the fundamental understanding of colloidal 2D semiconductor NPLs but also emphasize their significant potential for advancing the next generation of optical and quantum technologies.

AUTHOR INFORMATION

**Corresponding Authors**

*E-mail: michael.zopf@fkp.uni-hannover.de

*E-mail: jannika.lauth@uni-tuebingen.de

*E-mail: louis.biadala@iemn.fr

**Author Contributions**

J. L. and M. Z. conceived the project and supervised the experiments. Optical measurements were conducted by P. L. (low-temperature) and L. F. K. (room-temperature). The low-temperature measurements were supported by J. Y. and the respective data analysis by T. N. R., H. T. N. and A. A. performed the TEM experiments under the supervision of L. Biadala. I. Z. and F. S. were in charge of the X-ray experiments.

L. Biadala., J. L and M. Z. contributed strongly to the interpretation of the data. P. L. and E. P. R. prepared the samples for low temperature micro-PL. P. L. and R. G. performed the polarization simulation. The manuscript was written by P. L. and L. Biesterfeld., with input from all co-authors.

‡ These authors contributed equally.

**Notes**

The authors declare no competing financial interest.

ACKNOWLEDGMENT

The authors gratefully acknowledge the German Federal Ministry of Education and Research (BMBF) within the projects SemIQON (13N16291) and QVLS-iLabs: Dip-QT (03ZU1209DD), as well as the Deutsche Forschungsgemeinschaft (DFG, German Research Foundation) under Germany's Excellence Strategy (EXC-2123) Quantum Frontiers (390837967). P. L. acknowledges the China Scholarship Council (CSC201807040076). L. Biesterfeld., L. F. K. and J. L. gratefully acknowledge funding by the Deutsche Forschungsgemeinschaft (DFG, German Research Foundation) under Germany's Excellence Strategy within the Cluster of Excellence PhoenixD (EXC 2122, Project ID 390833453). J. L. is thankful for funding by the Ministry for Science and Culture of the State of Lower Saxony (MWK) for a Stay Inspired: European Excellence for Lower Saxony (Stay-3/22-7633/2022) Grant and for additional funding by an Athene Grant of the University of Tübingen (by the Federal Ministry of Education and Research (BMBF) and the Baden-Württemberg Ministry of Science as part of the Excellence Strategy of the German Federal and State Governments). L. Biadala. and H. T. N acknowledge the Agence National de la Recherche (Grant No. ANR-19-CE09-0022, "TROPICAL" Project), and the I-Site ULNE foundation - PEARL doctoral program (H2020 MSCA-COFUND-2018 GA n°847568, "METEOR" project)

The authors gratefully acknowledge Serguei Goupalov for fruitful discussions and Fei Ding for his strong support in the supervision and discussion of the experimental works.

The authors are grateful to Elena Chulanova, Anton Pylypenko, Ingrid Dax and Matthias Schwarzkopf for the help with GIWAXS measurements. We acknowledge DESY (Hamburg, Germany) for the provision of experimental facilities. Parts of this research were carried out at P03 beamline of PETRA III synchrotron (proposal I-20220914).

# Supplementary Notes

A. Methods

**Chemicals.** Acetonitrile (⩾ 99.5 %), cadmium(II) chloride (99.99 %) isopropanol (⩾ 99.5 %), lead(II) oxide (⩾ 99.99 %), methanol (⩾ 99.8 %), *n*-octylamine (99 %), oleic acid (⩾ 90 %), rhodamine 6G (~95 %), triethylamine (⩾ 99 %), trifluoroacetic acid (99 %), and trifluoracetic anhydride (⩾ 99 %) were purchased from Sigma-Aldrich/Merck. n-Hexane (99.99 %) and thiourea (99 %) were purchased from Alfa Aesar. Ethanol (99.9 %) was purchased from Acros. Tetrachloroethylene (⩾ 99.9 %) was purchased from Merck-Millipore. *n*-Octylamine and oleic acid were degassed *via* the freeze-pump-thaw method three times prior to being stored and used inside a nitrogen-filled glovebox. All other reagents were directly used as received from the listed suppliers without further purification.

All synthetic steps were performed under inert gas conditions inside a nitrogen-filled glovebox, unless explicitly stated otherwise.

**Preparation of the Lead Oleate Precursor.** Lead oleate was synthesized *via* an established method described by Hendricks *et al.*[1] For the PbS NPL synthesis, lead oleate (365 mg, 0.47 mmol) was weighed into a 8 ml screw cap vial and dissolved in a mixture of *n*-hexane (2.3 ml), *n*-octylamine (1.5 ml), and oleic acid (0.8 ml) by stirring at 35 °C until complete dissolution.

**Preparation of the Thiourea Precursor.** The thiourea precursor was prepared by dissolving thiourea (180 mg, 2.36 mmol) in n-octylamine (4.5 ml) under continuous stirring at 35 °C. The mixture was stirred for at least 30 h before being used in the PbS NPL synthesis.

**PbS NPL Synthesis.** The PbS NPLs were synthesized following a procedure by Manteiga Vázquez et al.[2] The 8 ml screw cap vial containing the premixed lead oleate solution was allowed to heat up to 35 °C and 0.5 ml of the thiourea precursor solution were rapidly injected. After a reaction time of 20 min the solution exhibited a bronze to dark-brown color and the PbS NPLs were passivated by adding a solution of $CdCl_2$ (2.5 ml, 0.1 M) in a mixture of *n*-octylamine and oleic acid (volume ratio of 9:1) and subsequent stirring for additional 10 min. The passivated PbS NPLs were stored at -25 °C in the fridge of a nitrogen-filled glovebox.

For preparing the samples for optical measurements at cryogenic temperatures the PbS NPLs were first transferred to toluene. Briefly, the NPLs where precipitated by drop-wise addition of a mixture of isopropanol and ethanol (3:1) until visible destabilization of the colloidal solution, centrifuged at 2500 rcf for 10 min, and reprecipitated in dry toluene.

The PLQY was determined using a relative method described by Würth et al.[3] using Rhodamin 6G in dry ethanol as a reference dye with a known quantum yield of 0.95. Thereby the PLQY is given by

$$PLQY = 0.95 \cdot \frac{A_{sample}}{A_{reference}} \cdot \frac{1 - 10^{-Abs_{reference}}}{1 - 10^{-Abs_{sample}}} \cdot \left(\frac{n_{TCE}}{n_{EtOH}}\right)^2 \text{, (S1)}$$

with the integral area *A* of the PL measurement curve, the absorbance *Abs*, and the refractive index *n* of the solvent.

**Transmission electron microscopy.** Overview TEM images were obtained using a FEI Tecnai G2 F20 transmission electron microscope equipped with a field emission gun operating at 200 kV. Samples for TEM analysis were prepared by drop-casting the colloidal PbS NPLs onto carbon-coated copper grids (300 mesh) acquired from Quantifoil.

**High-angle annular dark-field scanning transmission electron microscopy.** A FEI TitanThemis 300 microscope equipped with a probe aberration corrector, which is operated at 200 kV, was used to acquire (HR)STEM images. The probe size was set to 0.1 nm with a convergence semiangle of 22.5 mrad. The collection angle of the HAADF detector was in the range 80−150 mrad. The contrast in an HAADF image is proportional to $Z \approx 1.7-2$, meaning that the bright contrast indicates relatively heavy atomic composition.

**Grazing-Incidence Wide-Angle X-ray Scattering.** GIWAXS diffractions patterns were measured at the beamline P03 of the PETRA III synchrotron facility (DESY, Hamburg) at an incidence angle of $\alpha_i = 0.4°$. For this, samples were prepared by drop casting the colloidal PbS solutions onto silicon wafers (5 mm x 5 mm, p-typed doped with boron, <100> surface, purchased from Plano).

**Optical measurements at cryogenic temperature:** The optical measurements at cryogenic temperatures for the ensembles of PbS NPLs and the individual PbS NPLs were performed using a closed-cycle helium flow cryostat (Montana Instruments, Cryostation C2). The drop-casted PbS NPL samples were prepared the following way: 5% weight fraction of polystyrene were added to the PbS NPL solution in toluene, which was then centrifuged at 500 rpm for 2 min. The solution was then drop-casted onto a flat silicon substrate covered with a gold film, acting as mirror to increase the yield of collected PL from the sample. For obtaining the emission spectra of NPL ensembles and single NPLs, they were excited using continuous wave diode laser light at 532 nm (Thorlabs) which was focused onto the sample by an objective (Mitutoyo, M Plan Apo NIR 100x) with a numerical aperture of 0.7. The emitted PL signal was guided to a spectrometer (Spectroscopy & Imaging GmbH) with 300 l/mm grating and detected by a CCD (see figure S4). We calibrated the spectral transmission of our setup and spectrometer by guiding a broad-band white light signal (Thorlabs SLS202L) through the optical setup, including the spectrometer. The recorded spectrum and the original spectrum of the white light source were used to calibrate the total spectral efficiency of our setup and spectrometer. Different NPL ensemble emissions have been studied at different positions on the sample. For the time-resolved PL measurements, the PbS NPLs were excited by pulsed laser light at 445 nm for the ensemble of NPLs, and at 500 nm for the single NPLs. The repetition rate of the pulsed laser was 82 MHz. The signal was collected by an avalanche photodiode. For the time-resolved PL measurements on ensemble PbS NPLs, the data is fitted bi-exponentially:

$$I(t) = I_0 + A_1 e^{\frac{-t}{\tau_1}} + A_2 e^{\frac{-t}{\tau_2}}, \quad (S2)$$

With the background intensity $I_0$ and two amplitude components $A_1$ and $A_2$. These correspond to the intensities of the slow and fast decay components, with respective lifetimes $\tau_1$ and $\tau_2$, For the single PbS NPLs. the data is fitted single-exponentially:

$$I(t) = I_0 + A e^{\frac{-t}{\tau}}, \quad (S3)$$

where $I_0$ represents the background intensity and A denotes the amplitude and $\tau$ is the derived lifetime.

For the polarization-dependent measurements, the degree of polarization (DOP) was determined by $\delta = (I_{max} - I_{min}) / (I_{max} + I_{min})$, where $I_{max}$ and $I_{min}$ are the fitted maximum and minimum PL intensities, respectively.

**B. Polarization-dependent photoluminescence analysis.**

In this section, we present the theoretical model for the simulation of the polarization degree dependent on the NPL orientation on the substrate. To determine the orientation $(\Theta, \Phi)$ of an individual nano-emitter, we employ the theoretical framework proposed in Refence [4]. Our simulated scenario is as follows. As shown in [4], the degree of linear polarization of the emission is defined as:

$$\delta(\Theta) = \frac{I_{max} - I_{min}}{I_{max} + I_{min}}, \quad (S4)$$

where $I_{min}$ and $I_{min}$ are the minimum and maximum intensity. For 1D dipoles,

$$I_{min} = A\sin^2\Theta + B\cos^2\Theta, \quad (S5)$$

$$I_{max} = C\sin^2\Theta + I_{min}. \quad (S6)$$

And for 2D dipoles,

$$I_{min} = A + B + (A - B + C)\cos^2\Theta, \quad (S7)$$

$$I_{max} = C\sin^2\Theta + I_{min}. \quad (S8)$$

The constants A, B, and C can be determined analytically. Particularly the simple expression for δ in the limit of high numerical aperture ($\theta_{max} = \pi/2$) for a 1D and 2D dipoles are:

$$\delta_{high\,NA,1D}(\Theta) = \frac{7}{8}\sin^2\Theta, \quad (S9)$$

$$\delta_{high\,NA,2D}(\Theta) = \frac{7}{16}\sin^2\Theta, \quad (S10)$$

respectively. For both 1D and 2D dipoles the polarization degree δ depends on the out-of-plane Θ. Moreover, in the limit of low numerical aperture, δ for 1D and 2D dipoles[4] are:

$$\delta_{low\,NA,1D}(\Theta) = \frac{\sin^2\Theta}{(1 - ((\theta_{max}^2)/2)\sin^{\wedge}2\Theta + ((\theta_{max}^2)/2)}, \quad (S11)$$

$$\delta_{low\,NA,2D}(\Theta) = \frac{\sin^2\Theta}{\left(((\theta_{max}^2)/2) - 1\right)\sin^2\Theta + 2}, \quad (S12)$$

respectively. If we use low numerical aperture, i.e., NA = 0.7, In the case of $n_1$ = 1.5 (Polystyrene index), and $n_2$ = 1 (medium index), we can apply Eqn.(S12) to show δ change with Θ, with

$$\theta_{max} = \arcsin\frac{NA}{n_1}. \quad (S13)$$

The calculation result is presented in Figure S8.

**C. Phonon sidebands in the photoluminescence spectra**

All emissions that we observe feature sidebands which we attribute to phonon-assisted emission processes. Usually, a low energy tail of the PL is seen, which we attribute to the coupling with acoustic phonons. Coupling to distinct optical phonon related features seems to be not very prominent in the PbS NPLs. Such coupling is common in semiconductor nanocrystals and is typically explained by a mixture of surface-related and intrinsic effects[5]. The presence of localized surface-charges induces a polarization in the nanocrystal and therefore enhances coupling to longitudinal (LO) phonons. Due to the high crystal and surface quality and passivation in our PbS NPLs leading to little spectral diffusion and the presence of a narrow emission, this type of coupling is expected to be strongly reduced. Intrinsic effects include a reduced overlap of electron and hole wavefunctions in the NPL leading to an increased polarization in the nanocrystal and therefore enhanced coupling to LO phonons. In the case of PbS NPLs, the wavefunction overlap is maximized in the direction of strongest confinement, and the quality of the orthogonally oriented surface is high with negligible surface reconstruction. In contrast, coupling to acoustic phonons (as seen in the spectrum) is expected via deformation potential and piezoelectric interactions. Until now, this suppression of optical phonon modes has not been observed in PbS nanostructures, most likely because of strong internal dipoles caused by asymmetry in the polar facet orientation and the structure truncation. Due to the high crystal and surface quality with negligible surface reconstruction (see Figure 1) as well as the applied passivation in the 2D PbS NPLs, a high wavefunction overlap in the direction of strongest confinement is expected, and thereby strongly reducing the NPL polarization and LO phonon coupling. The observed low-frequency acoustic modes (both in Stokes and anti-Stokes PL) are most likely stemming from the thickness breathing mode.[6,7] The phonon energies are in the range of a few meV with slight changes from NPL to NPL (see Figure 4 and SI), which is reasonable for confined acoustic phonons that are more sensitive to symmetry and size. The observed phonon energy range aligns well with transient absorption spectroscopy results in similar (slightly smaller) PbS NPLs.[2]

# Supplementary Figures and tables

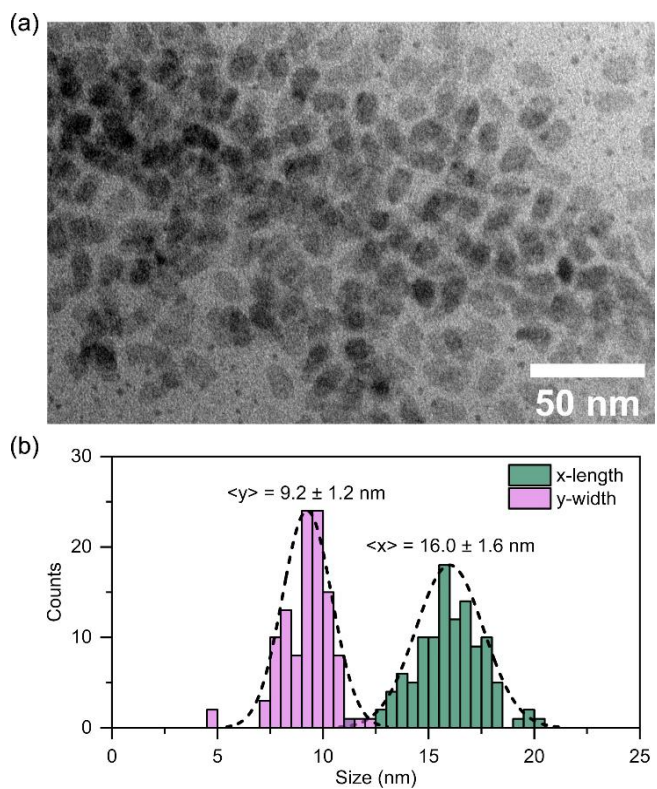

**Figure S1.** (a) Typical overview TEM image used for determining the lateral dimensions of the PbS NPLs. Within areas of higher concentration (compared to the overview image shown in the main manuscript), individual NPLs overlap with each other. (b) Corresponding size histogram, x-lengths correspond to the longest dimension of the NPLs, y-widths are the longest distance orthogonal to the x-length.

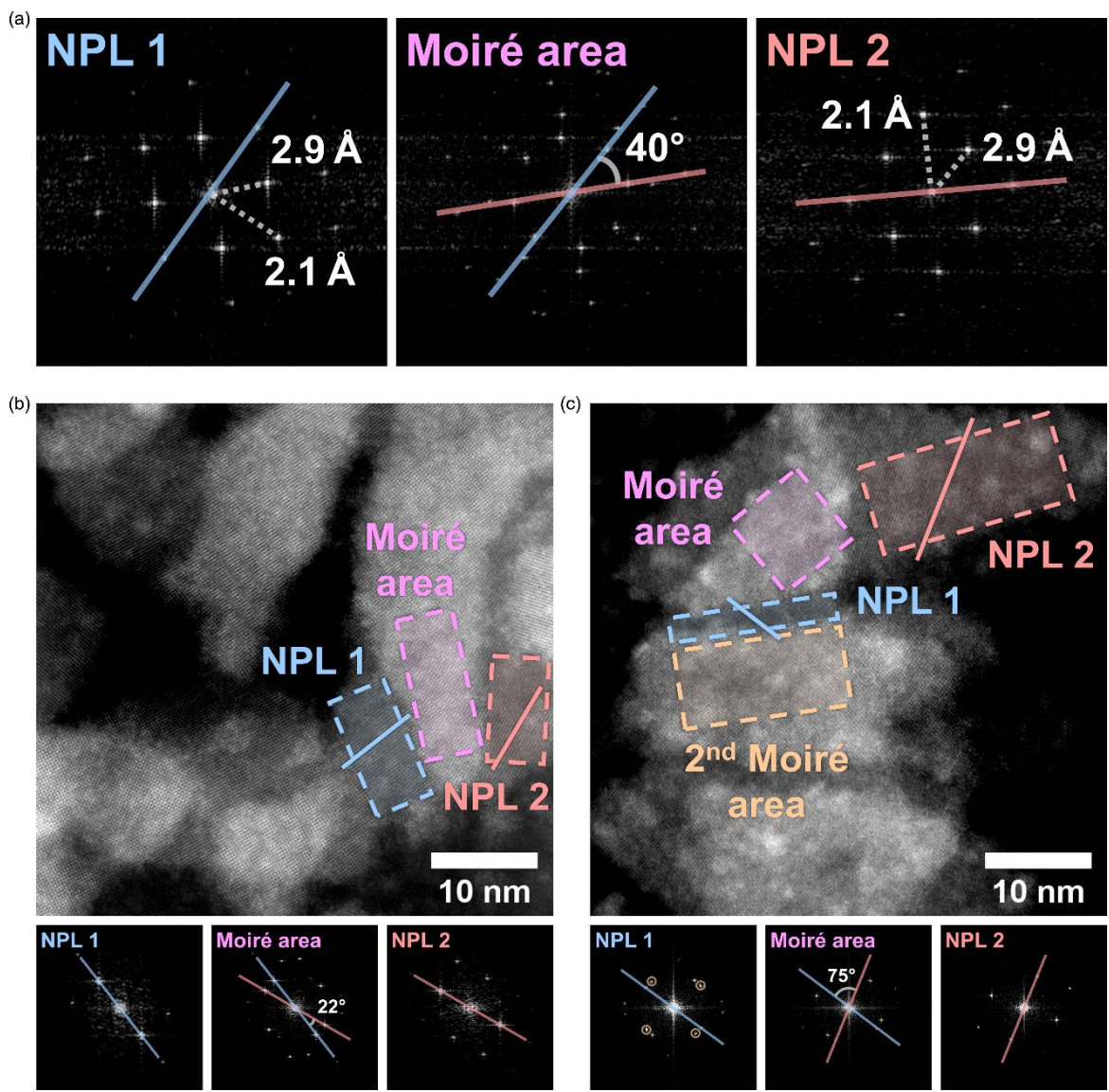

**Figure S2.** (a) FFT patterns corresponding to the two NPLs and the Moiré pattern depicted in Figure 1b of the main manuscript. The two PbS NPLs overlap with a twist angle of 40°. (b, c) HR-HAADF-STEM images and corresponding FFT patterns depicting additional examples of Moiré patterns formed by overlapping PbS NPLs with twist angles of (a) 22° and (b) 75°, respectively (compare to Figure 1b of the main manuscript).

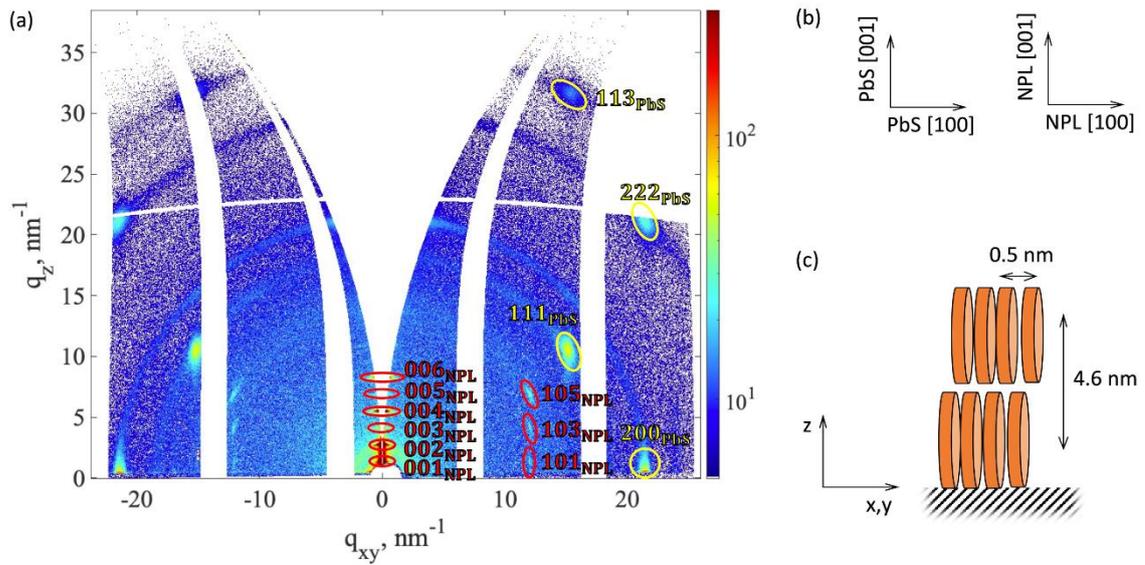

**Figure S3.** (a) Grazing-incidence wide-angle X-ray scattering diffraction pattern (background-corrected) of PbS NPLs drop-casted onto a silicon substrate (white areas are detector gaps). The presence of superlattice peaks (marked in red) at small values of the scattering vector $q$ indicate the formation of a superlattice of the individual NPLs. At the same time, the Bragg peaks from the PbS atomic lattice (highlighted in yellow) allow us to confirm that the NPLs are oriented in a certain way within the formed superlattice. (b) Orientation of the crystallographic directions of the PbS atomic lattice and the superlattice of the NPLs. (c) Combining the Bragg peaks with the superlattice peaks, we propose the shown superlattice structure with the characteristic spacings of 0.5 nm and 4.6 nm between the individual NPLs in real space. Interestingly, the low intensity of the $hkl$ peaks with $h + k + l =$ odd suggests, that the rows of NPLs might be shifted by half of a unit cell with respect to each other, resembling a body-centered superlattice. The small distance of approximately 0.5 nm between the individual NPLs eliminates the possibility of relatively large organic ligands to fit between the NPLs.

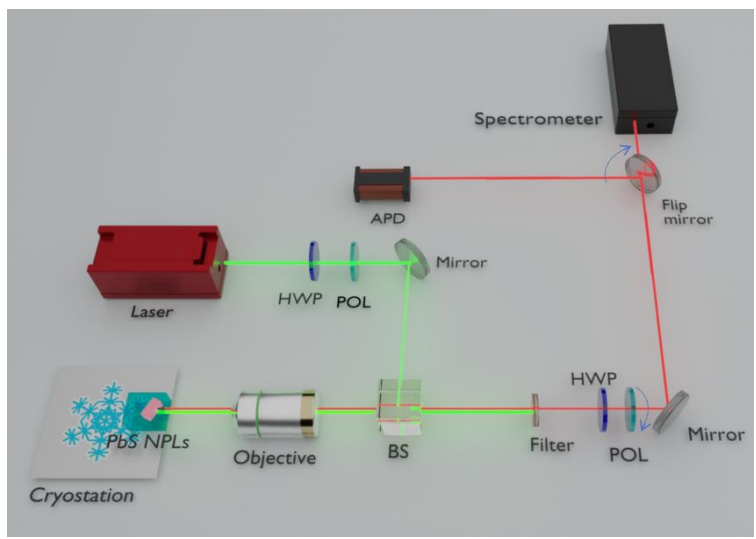

**Figure S4:** Sketch of the optical setup for micro-PL. BS: beam splitter. HWP: half wave plate. POL: polarizer. APD: Avalanche photodiode.

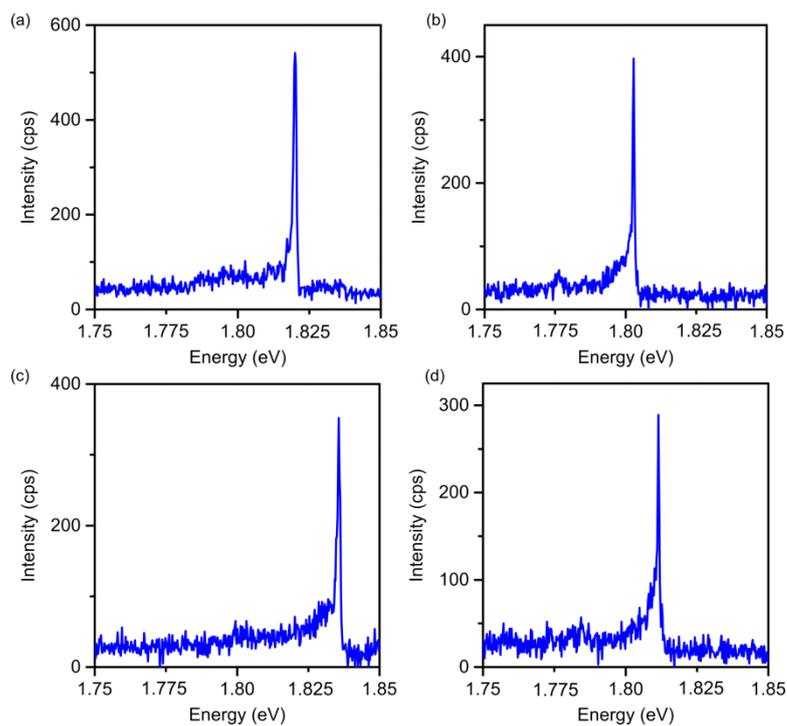

**Figure S5:** (a) – (d) Exemplary PL Spectra of four different single PbS NPLs at T = 4 K.

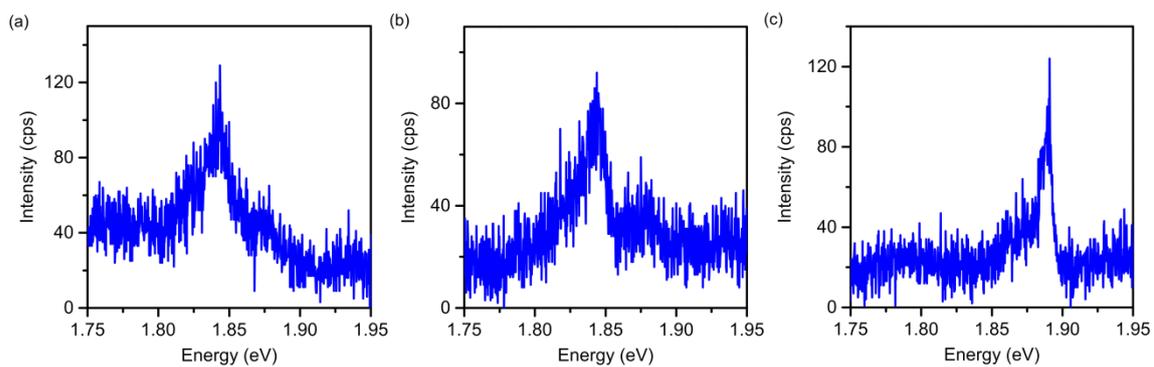

**Figure S6:** (a) – (c) Exemplary PL spectra (T = 4 K) of other localized emissions observed in the PbS NPL sample that do not exhibit narrowband emission.

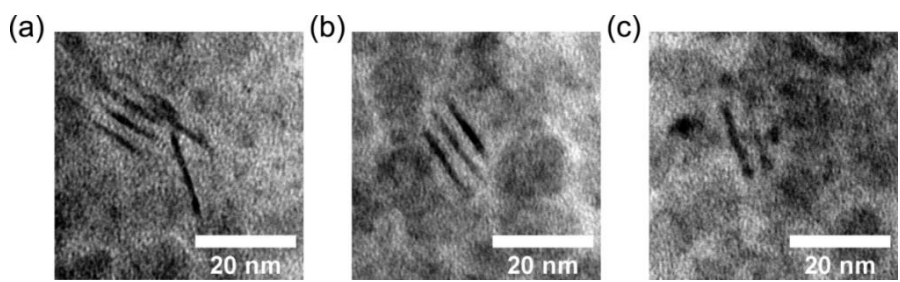

**Figure S7:** (a) – (c) Exemplary TEM images of PbS NPLs lying on their side exhibiting a thickness of 1-2nm.

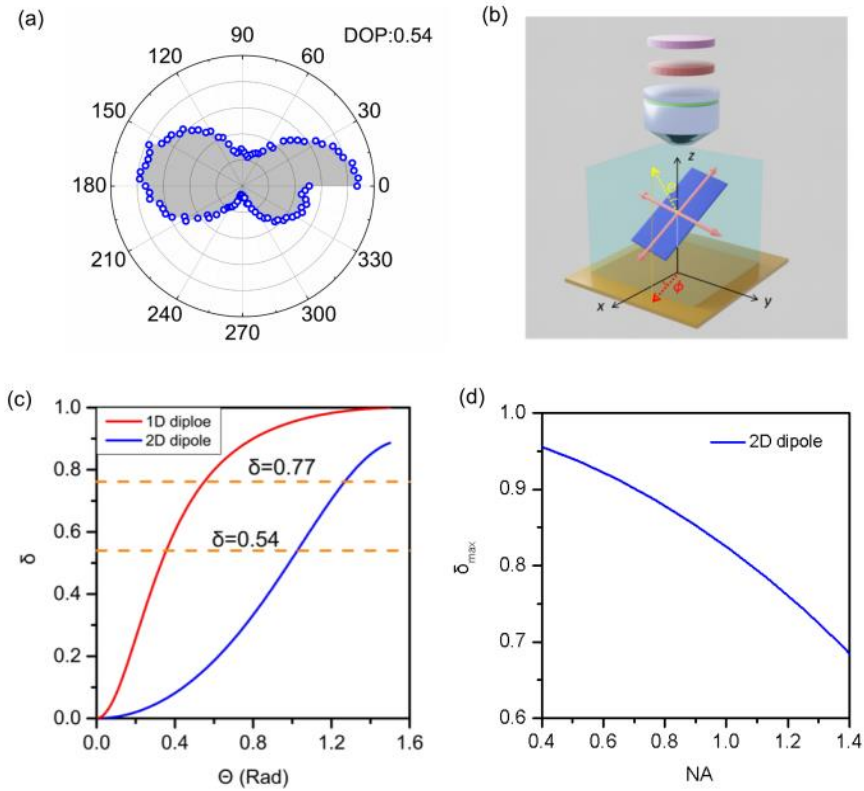

**Figure S8:** (a) Polarization angle dependent PL intensity of a single PbS NPL. (b) Sketch of the optical set-up and relevant parameters of the model for estimating the polarization degree of a single PbS NPL. (c) Calculated polarization degree over the collection angle for an ideal 1D dipole (red) or 2D dipole (blue), assuming the experimental setup in (b). (d) Calculated maximum degree of polarization as a function of the numerical aperture of the objective, assuming the experimental setup in (b)

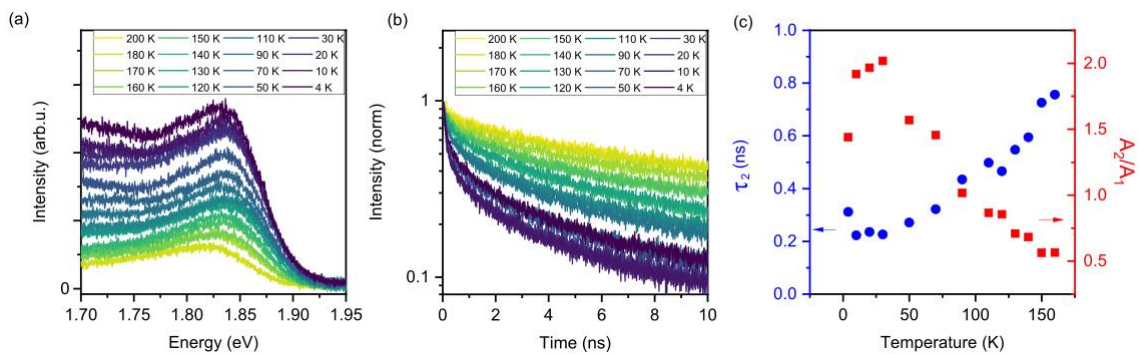

**Figure S9.** (a) Temperature-dependent PL spectra of the PbS NPL ensemble shown in Figure 3 of the main manuscript. (b) Fluorescence lifetime measurements of the ensemble of PbS NPLs. (c) Temperature-dependent decay time of the short component of the PL decay τ₂ (blue dot) and Temperature-dependent amplitude of short component of the PL decay over long component of the PL decay (red dot).

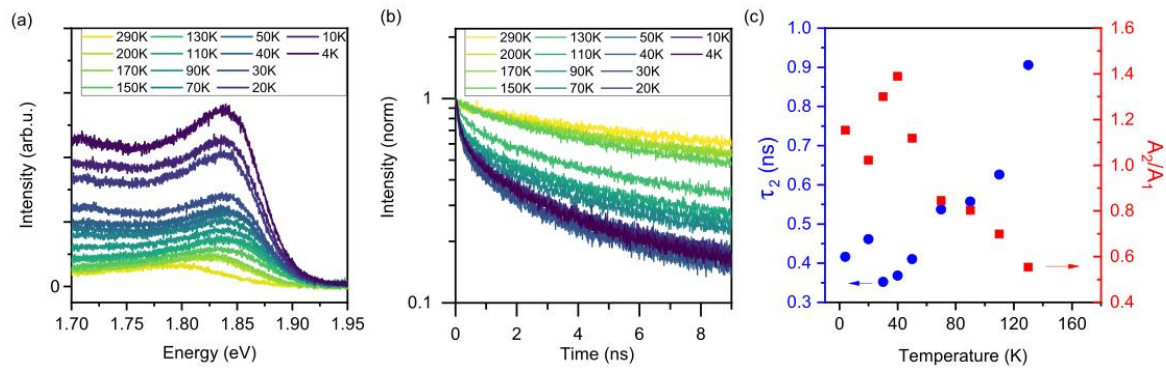

**Figure S10**. (a) Temperature-dependent PL spectra of a different ensemble of PbS NPLs. (b) Fluorescence lifetime measurements of the ensemble of PbS NPLs. (c) Temperature-dependent decay time of the short component of the PL decay $\tau_2$ (blue dot), the short decay component $\tau_2$ vanished at around 150 K and Temperature-dependent amplitude of short component of the PL decay over long component of the PL decay (red dot)

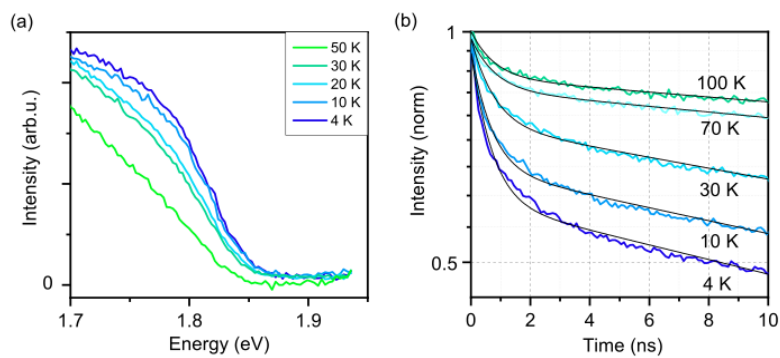

**Figure S11**. (a) Temperature-dependent, normalized PL spectra of a different ensemble of PbS NPLs. (b) Fluorescence lifetime measurements of the ensemble of PbS NPLs.

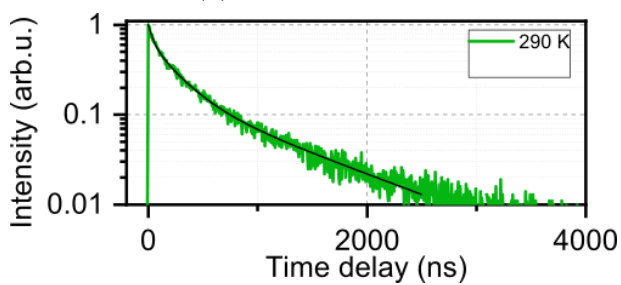

**Figure S12:** PL lifetime measurement of an PbS NPL ensemble at room temperature.

A triple-exponential decay model is employed, with $\tau_1$ = 42.6 ns, $\tau_2$ = 231.3 ns and $\tau_3$ = 956.2 ns.

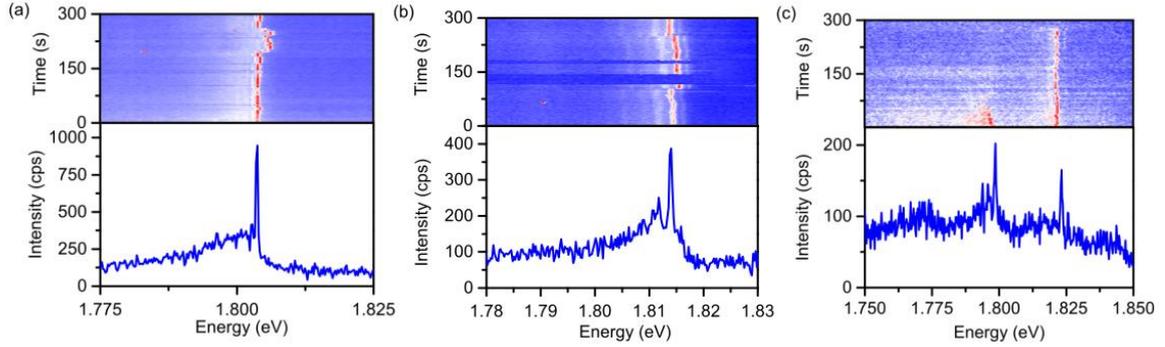

**Figure S13:** (a)-(c) PL time traces (top) and snapshot spectra (bottom) of three exemplary PbS NPLs exhibiting spectral diffusion.

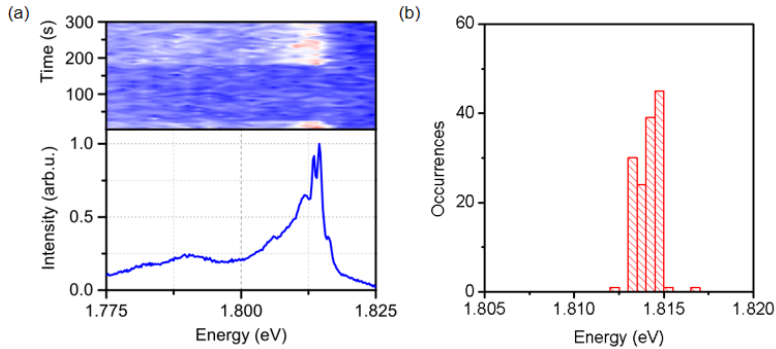

**Figure S14:** Only minor spectral diffusion and ultra-low blinking timescales in a representative PbS NPL. (a) The top part shows the PL time trace featuring low spectral diffusion and at the same time strong blinking behavior with a long off-time of around 150 s. The bottom part shows the normalized sum of spectra with the NPL in the bright state. (b) is the distribution of the central emission energy (obtained by fitting each spectrum in (a) with the NPL in its bright state).

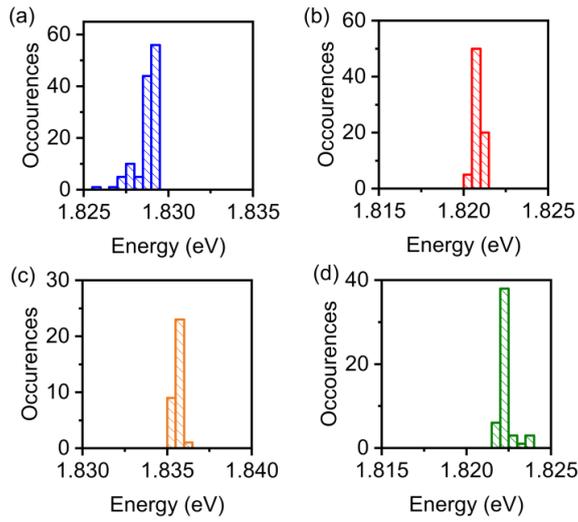

**Figure S15:** Distribution of the central emission energy of the strongest emission peak from four trion states in Figure 4e (obtained by fitting each spectrum in Figure 4e).

**Table S1:** The fitting results of specific PL emissions of PbS NPL derived from Figure 2a by using the Gaussian fittings.

|  | ZPL | Acoustic phonon |
|---|---|---|
| Position (eV) | 1.8028 | 1.7986 |
| FWHM (meV) | 0.615 | 7.250 |